\documentclass[pra,12pt,tightenlines,nofootinbib,floatfix]{revtex4} 
\pdfoutput=1
\usepackage{amsmath}
\usepackage{graphics, graphicx}
\usepackage{epsfig}
\usepackage{ulem}
\usepackage{color}
\usepackage{datetime} 

\begin{document} 




\begin{abstract}
 \centerline{ Brief recollections by the author about  how his  work with Kip Thorne influenced his career in physics.}
\end{abstract}

\title{Recollections of Kip Thorne}

\author{James  Hartle}
\affiliation{Department of Physics, University of California, Santa Barbara, California  93106, USA} {\affiliation{ Santa Fe Institute, \\ 1399 Hyde Park Road,  Santa Fe, New Mexico  87501, USA.}  

\bibliographystyle{unsrt}
\bibliography{references}


\maketitle

\section{Introduction}
\label{intro}


 My own modest  career  was significantly influenced  by working with Kip Thorne, by his guidance, 
  by his writings, and  by  interactions and discussions with him. This paper reports my memories of these interactions and influences.
  
   I am relying on my memory  for these bits of history knowing  all the risks that go along with that. I  cannot record what I have forgotten in the many years since the story began,  I am not writing accurate history that can be backed up by documents.  I am writing  about my recollections of situations and events that happened  long ago.
  
  I probably first met Kip at Princeton when he was  a graduate student of Johnny Wheeler and  I was a post-doc-lecturer there  having just completed my PhD  with Murray Gell-Mann at Caltech in 1963.   A few years later Kip and I  found ourselves not far apart in California---Kip as a professor at Caltech and me as a professor at the University of California, Santa Barbara.  We held occasional joint gatherings of Kip's group and mine; and these, under stimulus from Jim Isenberg, were transformed into the Pacific Coast Gravity Meetings.
  
    It happened in 1967 that Kip and I met at a summer workshop in Europe that we were both attending.   I discussed with him my work on deriving the explicit  equations of structure for slowly rotating relativistic stars. Kip proposed that when we both got back to California  we should solve these equations on a computer to produce realistic models of such general relativistic stars  and see what they told us. We did just that--- oscillating back and forth between  Pasadena, Santa Barbara and sometimes other places. This was my first exposure to any significant computing,  but not my last. 
  
  \section{The Kip Thorne School of  Computing}
  \label{kstschool} 
  Kip had definite ideas on how to organize computing:
  \begin{itemize}
  \item{}  Forget Runge-Kutta and just use the simplest trapezoidal rule.  
  \item{}Forget optimizing for expense and spend as much as is necessary to get the job done. 
  \item{} Stay up all night if you  have to, but get the job done by the morning.  (And Kip and I did stay up very late in the computer room at Caltech.) 
  \end{itemize} 
  
If memory serves,  Kip and I were using  an IBM 7094 at  Caltech and a similar machine at  UCSB.  Input was by punched cards and output was on many sheets of large computer paper.  The 7094's  computing power  is sometimes compared to that of a present day musical Japanese Christmas card. We thought of a program with three hundred Fortran statements as a {\it large} program.  We published our numerical models and their implications in  (J. B. Hartle  and K.S. Thorne, {\it Slowly Rotating Relativistic Stars, II. Models for Neutron and Supermassive Stars}, Ap. J. {\bf 153}, 807-834, (1968)).

  Thus began my effort in significant computing which hitherto I had avoided. It became  a significant part of my effort  at various times later but always reluctantly.  See e.g. (J. B. Hartle, {\it My Timeline in Quantum Mechanics}), forthcoming.
  
\section{Stability and Gravitational Waves from Pulsating, Slowly Rotating Stars}  

  To find the equations for calculating rotating, pulsating relativistic stars required expanding the Einstein  equations  to first order in the amplitude  of pulsation and second order in the rate of rotation. We did the job using  a then recently developed algebraic computer language called FORMAC.  (Kip always ahead of the pack.)  It was a tour-de-force of computation and, together with follow-on numerical computations, led to two papers on the influence of rotation on the stability of relativistic stars:
(J. B. Hartle  and K.S. Thorne, {\it Slowly Rotating Relativistic Stars III. Static Criterion for Stability}, Ap. J. , 158, 719-726, (1969));  also   (J. B. Hartle,  K.S. Thorne and S. M. Chitre, {\it Slowly Rotating Relativistic Stars VI. Stability of the Quasi-Radial Modes}, Ap. J. , 176, 177-194, (1972)). 

   With stability understood, we undertook a study of how the star's rotational flattening (an effect second order in the rotational angular velocity) causes its non-radiating radial pulsations to become quasiradial and thereby emit gravitational waves.  The equations for this were far more complex than for stability--a super tour-de-force for the FORMAC algebraic manipulation code of that era.  We completed the computer-aided analytic analysis, drafted a paper, and then embarked on a multi-decade shuttling of the manuscript back and forth between Santa Barbara and Pasadena.  As I write this Kip and I are entering the fiftieth year of massaging this manuscript,  (J. B. Hartle,  K.S. Thorne, {\it Slowly Rotating Relativistic Stars VII: Gravitational Radiation from the Quasi-Radial Modes},  unpublished ).  Perhaps now would be a good time to bring it to completion, and perhaps our lack of diligence in completing this work may be an object lesson to our younger colleagues.

\section{Equations of Motion} 
\label{eqmot} 
The force on a charged particle that defines its motion in an electromagnetic field is a separate postulate from Maxwell's equations.  But in general relativity the motion of particles  in curved spacetime follows from  the gravitational field equations. This was first shown by Einstein, Infeld and Hoffmann, and  many others since.  Kip and I showed it yet again, though this time for black holes, in 
K.S. Thorne and J.B. Hartle, {\it Laws of Motion and Precession for Black Holes and Other Bodies,} Phys. Rev. D {\bf 31}, 1815-1837 (1985). The extension to precession was important because some black holes had  spiraling jets that could be caused by precession. 

My contributions to the equations of motion paper were important but  modest compared  to those of Kip.  I insisted that Kip and I appear as authors in anti-alphabetical order to recognize this. At first Kip refused but eventually he was persuaded.


\section{Later}
\label{later}  
After the equations of motion paper Kip and I moved in different directions: Kip's research  was focussed on the genesis and detection of gravitational waves i.e. towards LIGO.  I moved in the direction of quantum mechanics and cosmology  --- working with people like Murray Gell-Mann and Stephen Hawking on the wave function of the  Universe. 

:   
\section{Conclusion} 
\label{conclu}
Were I asked to describe Kip Thorne in two  words  I would choose `a leader'. He manages to inspire and bring out the best in students, in postdoctoral  researchers, in  theorists, in experimentalists, in funding agencies, and even in scientists like me. Would we have LIGO with its Nobel  Prizes if not for Kip?  Would I have worked in general relativistic astrophysics if not for Kip?  Is it an accident that so many of today's  prominent workers in  general relativistic astrophysics   were his students and postdocs?. He is working  in a field he 
 helped build
and was a mentor to many now in it. 

\section{Acknowledgements} 
\vskip .1in 
\noindent{\bf Acknowledgments:}    Thanks  are due to the NSF for supporting its preparation under grant PHY-18-8018105 and to Mary Jo Hartle for proofreading it more than once  and to Kip for helpful suggestions.

 \end{document}